# Coalescence of Impurity Strontium Atoms in Lightly-Substituted Samples of Lanthanum Cuprate

*E. Yu. Beliayev[1], Dianela Osorio [2,3] and P. Contreras[3]*

[1] B. Verkin Institute for Low Temperature Physics and Engineering, National Academy of Sciences of Ukraine, Kharkiv, Ukraine.

[2] Department of Brain and Behavioral Sciences, University of Pavia, Pavia, Italy.

[3] Department of Physics, University of the Andes, Mérida, Venezuela.

**Abstract:** By comparing experimental results obtained in the study of the electron transport and magnetic properties of samples of lightly strontium-substituted lanthanum cuprates and the results of a previously published numerical analysis in the reduced elastic scattering space (RPS) of the elastic scattering cross-section, we construct a phenomenological classical self-consistent strong-binding energy density profile in the superconducting $La_{2-x}Sr_xCuO_4$ cuprate. It is found that the imaginary part of the RPS can be zero, almost constant, or strongly self-consistent, depending on the Sr substituted concentration. If the Sr atomic concentration is dilute, a constant imaginary part of the elastic scattering cross-section brings a sub-melted coalescent superconducting metallic profile similar to the one with a constant scattering lifetime in the metallic state of normal metals, except for the unitary narrow, sharp peak at zero frequency. A self-consistent broadening of this peak occurs for a higher concentration of Sr atoms, and finally, for very small amounts of non-magnetic dirt, only the narrow unitary peak prevails around the zero frequency.

**Keywords:** Coalescence, unitary limit, reduced elastic scattering phase space, line nodes, lightly-substituted lanthanum cuprates.

## 1.    Introduction.

This work is dedicated to understanding at experimental and phenomenological levels some classical properties in the unusual low-temperature behavior of unconventional, lightly-strontium-substituted lanthanum cuprate samples ($La_{2-x}Sr_xCuO_4$). We compare experimentally obtained evidence for the lightly-doped superconductor with a phenomenological analysis in the reduced elastic scattering phase space (RPS) to understand the relation between classical and quantum physical phenomena in lightly-doped cuprates.

The RPS is a physical self-consistent elastic scattering space based on the imaginary and real parts of the elastic scattering cross-section SCCS given by the pair of coordinates $(\Re(\tilde{\omega}), \Im(\tilde{\omega}))$, i.e., energy is conserved in the presence of an external non-magnetic potential $U_0$, but momentum "$k$" is not conserved. This happens if the fermionic mean-free path "$\ell$" is comparable to the inverse Fermi momentum "$k_F^{-1}$", or to the lattice parameter "a" following the physical kinetics relation $\ell\, k_F \sim \ell\, a^{-1} \sim 1$, which corresponds to the unitary limit ($c = 0$), where c is an inverse parameter to the external potential $U_0$ firstly proposed by Pethick and Pines [1] to explain low temperature transport properties in unconventional superconductors. The RPS resembles the *"phase space distribution probability in quantum-mechanics"* introduced by Wigner in 1932 [2].

In references [2, 3], the main feature of this approach is pointed out: *"the method is useful in providing easy reductions from quantum theory to classical physics kinetic regimes under suitable conditions"*. Therefore, in this work, we use this method to compare the interval $(\Re(\tilde{\omega}), \Im(\tilde{\omega}))$ as a function of the doping "$x_{theory}$" with previously reported evidence of experimental doping "$x_{exp}$" for some low concentration phases in strontium-substituted lanthanum cuprate samples on the threshold between the superconducting phase of the lightly-doped $La_{2-x}Sr_xCuO_4$ compound and antiferromagnetic phase of the Mott-Hubbard dielectric compound $La_2CuO_4$. With this tool, we explain part of Dalakova and co-workers' experimental findings [5 – 17].

This work is organized as follows. The section 2 shows the experimental results for low strontium doping in $La_{2-x}Sr_xCuO_4$, remarking some physical insights due to the strontium substitution in the original Mott-Hubbard dielectric. The section 3 proposes an energy self-consistent density profile without entering in details of how to derive the equations (previously explained and derived in [18,19]), sketching only the results for the SCCS $\Re(\tilde{\omega})$ and $\Im(\tilde{\omega})$ parts. The self-consistent phenomenological density energy profile $n(\tilde{\omega})$, is based on the unitary limit analysis of a 2D tight-binding [20] $La_{2-x}Sr_xCuO_4$ Scalapino nodal line model [21]. The $n(\tilde{\omega})$ analysis helps phenomenologically explain the origin of three phases for the Sr substituted HTSC compound. Finally, the Section 4, elaborates conclusions and recommendations, particularly on observations of the scattering lifetime and theoretical fits of the metallic coalescence state for unconventional superconductors.

## 2. Experimental low-temperature pieces of evidence for coalescence in lightly-doped $La_{2-x}Sr_xCuO_4$ samples with strong AFM correlations.

The experimental material presented in this article is based on a review of already published works by N. Dalakova and E. Beliayev (who is also one of the coauthors of this work) and their colleagues [5 – 7, 11 – 16]. For those studies, at St. Petersburg State University (SPbSU), a series of $La_{2-x}Sr_xCuO_4$ samples with strontium content $x = 0.001, 0.002, 0.005, 0.010$ were prepared by the solid-phase synthesis reaction by N. Bobrysheva and A. Selyutin in their laboratory. Synthesis details and the results of structural studies, which were carried out at St. Petersburg State University using SEM Cam Scan electron microscope, EDS LINK AN-10000, and WDS MIKROSPEC spectrometers, are given in [5– 7, 16]. The strontium content was determined in five areas for each sample obtained using a high-resolution WDS spectrometer. The magnetic properties (by PPMS-5M) and electron transport properties (using the homemade helium cryostat with Kapitsa-design electromagnet and measurement system based on Keithley 2182 nanovoltmeter and Keithley 2000 multimeters) were carried out at B. Verkin Institute for Low Temperature Physics and Engineering of the National Academy of Sciences of Ukraine.

For all the samples studied, the electron conduction properties were essentially nonlinear up to the appearance of $N$-shaped Voltage-Current Characteristics (VCC) [6, 14]. This assumed a strongly inhomogeneous phase state of the samples under study. Moreover, the degree of nonlinearity of VCC increased with a decrease in the degree of doping level $x_{exp}$.

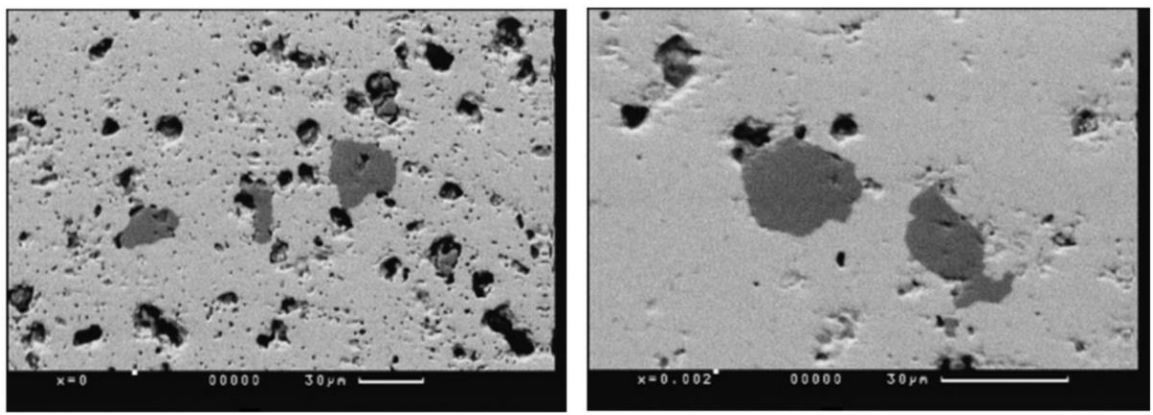

Fig. 1. Micrographs of thin sections of $La_{2-x}Sr_xCuO_4$ samples with $x = 0.001$ (left) and $x = 0.002$ (right) [5]

Additional XRD, SEM, and microprobe studies confirmed that despite all the efforts of the technologists, the samples turned out to be inhomogeneous on a microscopic scale. Microprobe analysis invariably showed that in the mixture with a strontium concentration of $x = 0.001$ after completing the solid-state synthesis, there were areas with either a Sr concentration of $x_{exp} = 0.002$ or a Sr concentration being close to zero (Fig. 1) [5]. For all other concentrations, there were no such problems. For Sr concentrations of $x_{exp} = 0.005$ or $x_{exp} = 0.01$, the strontium concentration in any sample part corresponds to the prescribed one. However, at very low $x_{exp} = 0.001$, the sample was prone to phase separation into two phases with almost zero strontium content and a strontium concentration close to $x_{exp} = 0.002$.

Also, the Neel temperature values $T_N$, measured for all the prepared low-substituted $La_{2-x}Sr_xCuO_4$ samples, which were supposed to increase monotonically with decreasing strontium concentration according to the known study by B. Keimer and co-workers [8] at the concentration of $x_{exp} = 0.001$, suddenly dropped sharply (Fig. 2 taken from [6] for this article has been supplemented by data taken from [5] at $x_{exp} = 0.002$, and by the commonly known $T_N$ value for stoichiometric $La_2CuO_4$). Fig. 2 demonstrates that free carriers (holes) that appear upon strontium doping level $x_{exp} = 0.001$ destroyed the strict AFM ordering in $La_2CuO_4$ Mott-Hubbard dielectric, leading to a sharp drop in $T_N$ values and severe deterioration in its antiferromagnetic characteristics. All the lines in Fig. 2 are only guides to the eye.

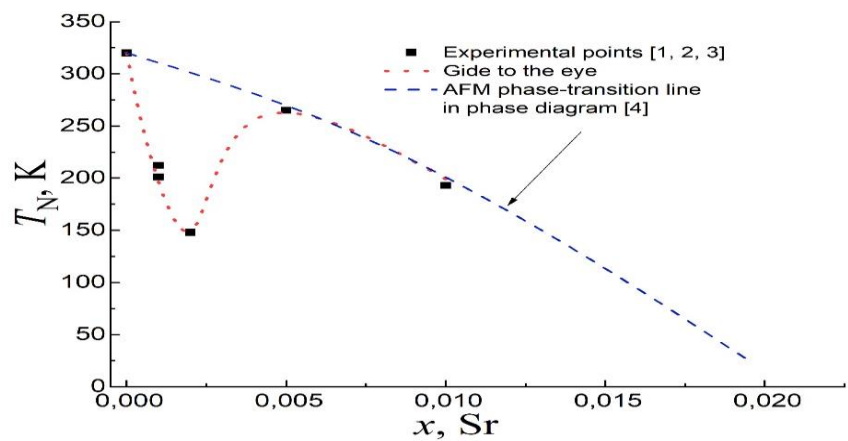

Fig. 2: Dependence of Neel temperature $T_N$ in strontium-substituted $La_{2-x}Sr_xCuO_4$ samples on Sr concentration. Lines are guides to the eye.

The coalescence phenomenon can also be explained using another physical language based on the "hot centers" picture [9, 10]. The substitution of a heavy La for Sr atom – the atom with a smaller mass but having a much larger interaction potential with neighboring atoms in the crystal lattice [9, 10], led to an increase in the frequency of local vibrations near a single localized strontium atom, thus significantly increasing the local temperature in that part of the crystal lattice. However, with an increase in the degree of doping, some beats have inevitably arisen between such hot impurity centers that disrupt the high-frequency resonance vibrations.

For this reason, at a specific concentration, the accumulation of strontium atoms in one place turns out to be energetically favorable since it lowers the local temperature. This is the physical basis for the coalescence phenomenon when single strontium atoms tend to merge into clusters. Having approached close enough distance, the "hot centers" began to interact through harmonics and lowered their energy of resonant vibrations, giving it to the lattice.

## 3. Numerical self-consistent evidence for coalescence in the unitary scattering limit for $La_{2-x}Sr_xCuO_4$.

In this section, we numerically propose a coalescence self-consistent energy density function profile $n(\tilde{\omega})$ with the help of a computational model based on the SCCS analysis for dressed normal quasiparticle carriers in strontium-substituted lanthanum cuprate already developed theoretically in our previous works [18, 19], aiming at comparing the results from this section using a dimensionless $x_{theory}$ concentration parameter defined by the relation

$$x_{theory} = \Gamma^+/\Delta_0 \quad (1),$$

where $\Gamma^+$ (that will be defined below) is the concentration parameter in the RPS space in meV, and $\Delta_0$ is the zero temperature ARPES value in meV of the superconducting gap. The aim of this section is to compare $x_{theory}$ with the previous experimental section analysis for the strontium substituted concentration $x_{exp}$ using the RPS SCCS method.

The reduced scattering phase space (RPS) with the pair of coordinates ($\Re(\tilde{\omega})$, $\Im(\tilde{\omega})$) is built with several parameters. The first group of three TB anisotropic parameters for a particular model: the superconducting gap at zero temperature $\Delta_0 = 33.9$ meV, the Fermi energy $\epsilon_F = 0.4$ meV, and the first neighbour hoping $t = 0.2$ meV, accordingly to the experimental ARPES data [26], using the Scalapino linear model [21, 22]. The normal state TB first neighbors energy is given by the expression $\xi(k_x, k_y) = \epsilon_F + 2t[\cos(k_x a) + \cos(k_y a)]$, with electron-hole symmetry $\xi(k_x, k_y) = \xi(-k_x, -k_y)$. This is the expression for a band centred at the corners ($\pm\pi/a$, $\pm\pi/a$) of the first Brillouin zone as is sketched in Fig. 3, where the nodal lines are the straight belts drawn in brown color and the anisotropic FS in clear blue color, using an implicit plot for the normal state energy and the singlet gap.

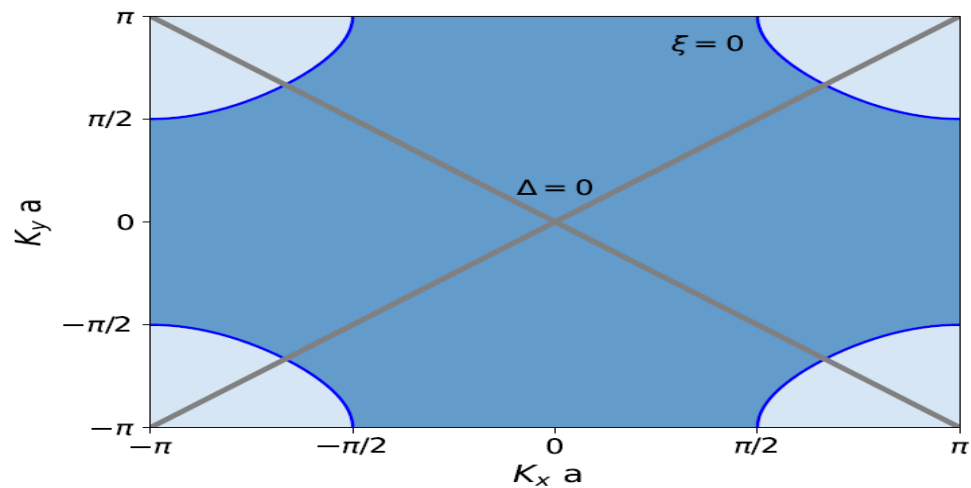

Figure. 3: The Scalapino linear nodal model for singlet HTSC such as strontium substituted $La_{2-x}Sr_xCuO$, showing in the brown lines the behavior of the nodal lines implicit plot, and in clear blue the tight-binding FS implicit plot.

In references [23, 24], it was theoretically and experimentally found that $La_{2-x}Sr_xCuO_4$ is in the unitary strong scattering limit if the Fermi surface is isotropic. They employed a specific superconducting electronic heat analysis at very low temperatures. The constant pressure electronic specific heat was measured by a conventional adiabatic heat pulse from 2 K to 10 K in order to extract the dimensionless residual density of states "$N(0)/N_F$" which is proportional to the superconducting specific electronic heat jump "$\gamma_s/\gamma_n$". By comparing the relations "$\gamma_s/\gamma_n \sim N(0)/N_F$ where hold the relation $c_{p-elect} = \gamma_s T$ at very low temperatures, using the equation for suppression of $T_c$ due to non-magnetic disorder, proposed by Larkin [27]. Using this procedure, Momono and Ido [23] extracted the residual DOS from very low temperatures electronic specific heat measurements at constant pressure, concluding that $La_{2-x}Sr_xCuO_4$ is in the unitary strong-scattering limit (c=0).

The original idea of parametrizing the SCCS based on the strength $c$ and the impurity concentration $\Gamma^+$ for isotropic HTSC with line nodes was taken from the work by Schachinger and Carbotte [25]. Additionally, in [25], the disorder parameters for non-magnetic parameters are defined as follows: for the nonmagnetic impurity concentration $\Gamma^+ = n_{imp} (\pi^2 N_F)^{-1}$, and the atomic nonmagnetic Sr atom strength $c = (\pi N_F U_0)^{-1}$.

This allowed us to build a complicated numerical procedure where we linked the TB parameters taken from Arpes measurements with the two nonmagnetic disorder parameters in a self-consistent way for very low frequencies needed to properly track the unitary regimen of the elastic SCCS, which is defined as

$$\widetilde{\omega}(\omega + i\, 0^+) = \omega + i\, \pi\, \Gamma^+ \frac{1}{g(\widetilde{\omega})},$$

and $g(\widetilde{\omega}) = \langle \frac{\widetilde{\omega}}{\sqrt{\widetilde{\omega}^2 - \Delta^2(k_x, k_y)}} \rangle_{FS}$ is a tight-binding numerical averaged. The whole set of microscopic parameters that can numerically be controlled in the fittings are five in this case, namely $(\epsilon_F, t, \Delta_0, c, \text{ and } \Gamma^+)$.

Furthermore, following Scalapino linear nodal idea, the first neighbours' superconducting gap terms in the irreducible tight-binding representation $B_{1g}$ corresponds to an even paired anisotropic singlet order parameter basis function $\phi(k_x, k_y) = \phi(-k_x, -k_y) = \cos(k_x a) - \cos(k_y a)$, with the gap equation $\Delta(k_x, k_y) = \Delta_0 \phi(k_x, k_y) = \Delta_0 [\cos(k_x a) - \cos(k_y a)]$. The superconducting gap for this symmetry corresponds to the one-dimensional irreducible representation $B_{1g}$ of the tetragonal point symmetry group $D_{4h}$ (as seen in Fig. 3). In this model, non-magnetic disorder induced by the atomic external strontium potential "$U_0$", strongly quenches superconducting ordering leading to suppression of $T_c$. The Ginzburg-Landau coefficients correspond in this case to (1,0).

Based on Fig. 4 and the theory and calculations elaborated in references [18, 19], we propose the empirical relation (2) following the classical theory of active collisions in physical kinetics, but taking into account, that the physical parameter $L$ of the problem is proportional to the lattice constant $a$ (unitary limit condition), i.e., $L \sim a$, and for the coalescence metallic region we notice that $\Im[\sigma] \sim \Im[\widetilde{\omega}] \sim (n_{dressed}(\widetilde{\omega}) - \delta \Im(\widetilde{\omega}))/(a\, n_{condensate})$ where a is the lattice parameter, $n_{dressed}$ refers to fermion concentration, and $n_{condensate}$ to boson concentration (readers interested in those details can follow the aforementioned works and references therein).

Aiming at helping to understand the coalescence region observed by N. Dalakova and co-workers in a series of experimental works [5-8] introduced and explained in Section 2, the function which we call the energy density self-consistent profile $n_{dressed}(\widetilde{\omega}) \sim \Gamma^+ [g(\widetilde{\omega})]^{-1} \sim \tau_{sup}^{-1}$ is proportional to the imaginary part of the self-consistent elastic scattering cross-section $\Im[\sigma]$, (in rationalized Planck units, it is given by meV and is suitable for an experimental contrast with the theory). The classical analysis elaborated for the SCSC can qualitatively separate three different phases based on a dimensionless concentration parameter $x_{theory}$ in the following way:

$$n_{dressed}(\widetilde{\omega}) \sim \begin{cases} \to 0, & \Gamma^+ = 0.01\, meV\ (\text{Mott Hubbard dielectric}) \\ constant + \delta\, \Im(\widetilde{\omega}), & \Gamma^+ = 0.05\, meV\ (b - \text{lighty subtituted Coalesc. } La_{2-x}Sr_xCuO_4) \\ \Im(\widetilde{\omega}), & \Gamma^+ > 0.05\, meV\ (c - \text{strange SC metal } La_{2-x}Sr_xCuO_4) \end{cases} \quad (2)$$

In relation (2), we can distinguish and observe three phases that can be compared with the experimental results from Section 2 where the most important feature observed in this unitary computational model, that attracted our attention when we discussed our findings with the experimental Kharkiv group was that at very low concentration $\Gamma^+$, only the central unitary resonance peak at zero frequencies seems to remain, i.e., the imaginary part of the elastic scattering cross-section $\Im(\widetilde{\omega}) \xrightarrow{yields} 0$ and $\delta\, \Im(\widetilde{\omega}) \ll 1$, for most of the superconducting frequency interval, that in the language of physical kinetics means that there are no dressed quasiparticles outside this narrow zero frequency interval, i.e., their inverse scattering lifetime $\tau^{-1} \xrightarrow{yields} 0\, meV$ (see Fig 4, left region shadowed very light blue color).

Therefore, it is proposed in this section that this resonance could be a signature of a phase transition to the antiferromagnetic phase of the Mott Hubbard dielectric $La_2CuO_4$ as Sr non-magnetic atoms concentration is reduced to zero since $x_{theory} \xrightarrow{yields} 0$.

It is worth noticing that in a previous work [28] (see Fig. 4, the yellow line in that paper), we observed that in the compound $Sr_2RuO_4$ at a very low concentration, $\Gamma^+$, the same resonance persists for the unitary strong scattering regime ($c = 0$). In [28], the computational calculation of the SCCS gives a larger mean averaged width of the imaginary part, nevertheless limiting the reduced scattering space to a small energy peak interval around the zero energy superconducting frequency.

In order to conclude this short but instructive numerical section analysis, we give Table 1 based on the calculation using the anisotropic TB parameters in agreement with the experiments mentioned and the posteriorly numerical calculation:

Table 1: Phases as a function of doping for antiferromagnetic $La_2CuO_4$ and lightly-strontium substituted metallic samples of $La_{2-x}Sr_xCuO_4$ where the doping is given by $x_{theory} = \Gamma^+/\Delta_0$ and $x_{exp}$

| Phases | a - MH dielectric $La_2CuO_4$ | b- Lightly-doped coalescent $La_{2-x}Sr_xCuO_4$ | c - Strange metal phase | idem | Idem |
|---|---|---|---|---|---|
| $\Gamma^+$ (meV) | 0.01 | 0.05 | 0.10 | 0.15 | 0.20 |
| $x_{theory}$ | 0.0003 | 0.002 | 0.003 | 0.004 | 0.010 |
| $x_{exp}$ | 0.001 | 0.002 | - | 0.005 | 0.010 |

A detailed analysis of Table 1 and Fig. 4 is as follows. Region (a) shadowed very clear blue in the left part of Fig. 4 represents an almost zero imaginary self-consistent scattering cross-section $\Im(\widetilde{\omega}) \sim 1/\tau(\widetilde{\omega}) \to 0$ for most of the entire energy window interval $\omega \in (-40, +40)$ meV. It means there are no dressed by non-magnetic impurity scattering (due to a $U_0$ potential) quasiparticles, except a few at the unitary resonance region around the zero frequency ($\Gamma^+ = 0.01$ meV $\to x_{theory} = 0.0003 \sim x_{exp}$).

We attribute this region to being a signature of the Mott-Hubbard dielectric $La_2CuO_4$ antiferromagnetic phase, which N. Dalakova, E. Beliayev and co-workers reported in their experimental studies on the results of strontium-substituted $x_{exp}$ doping parameter as being in essence $La_2CuO_4$. In the previous numerical work with the same model, we reported this phase as the one where $\tau^{-1}(\widetilde{\omega}) \to const \ll 1$, without comparing with experimental data since we did not have it at that moment [19].

Region (b) in the left part of Fig. 3 is given by the clear blue curve ($\Gamma^+ = 0.05$ meV). It represents the almost constant imaginary scattering cross-section $\Im(\widetilde{\omega}) \sim \tau^{-1}(\widetilde{\omega}) \to constant$, for most of the entire ω interval with a small amount of dressed by non-magnetic impurity scattering quasiparticles at the impurity concentration given by the slightly higher value of the concentration parameter $\Gamma^+ = 0.05$ meV except for a small resonance $\delta \Im(\widetilde{\omega})$. It corresponds to $x_{theory} = 0.002$ following the empirical relation (1) and this represents the metallic coalescing phase for a lightly doped $La_{2-x}Sr_xCuO_4$.

To explain coalescence, we propose that since linear momentum "***k***" is not conserved in the unitary limit for the elastic scattering cross-section, it is proposed that the dressed quasiparticles' momentum ***k*** is transferred to strontium atoms in the crystal lattice. Those Sr atoms migrate through the lattice and stick together in a coalescing metallic state, but only for a minimal concentration of Sr atoms, as was observed experimentally by Dalakova, Beliayev and co-workers in their results. Therefore, near the superconducting-antiferromagnetic transition in lightly strontium-substituted lanthanum cuprate, a robust metallic coalescing phase with an almost constant scattering lifetime can be interpreted theoretically and experimentally from a classical point of view and it is supported by the theory of active collisions in physical kinetics.

Region (c) represents, what we call the strong self-consistent imaginary phase of the SCCS for most of the entire frequency region when $\Im(\widetilde{\omega}) \sim \tau^{-1}(\widetilde{\omega})$, and $\Gamma^+ = (0.05, 0.10, 0,20)$ meV. It is observed for the values of the empirical relation (1) $x_{theory} = (0.003, 0.004, 0.010)$, shown in Fig. 4 shadowed gray. Undoubtedly, the value $\Gamma^+ = 0.20$ meV and $x_{theory} = 0.010$. It represents the so-called in the current literature strange metal phase in cuprates, where RPS self-consistent effects in the imaginary part of the SCCS mix the nodal Fermi dressed quasiparticles and superconducting pairs, making it very difficult to explain the transport properties based in parameters such as the mean lifetime "$\tau$" and free path "$\ell$" with strongly self-consistent frequency-dependency.

From the point of view of the experimentally observed electronic transport properties of lightly-doped lanthanum cuprates, attention may be drawn to Fig. 1. There, one can see a resonance at near-zero $x_{exp}$ concentration. Thus, the hypothesis arises that it is precisely this resonance that can explain the fact that in a series of aforementioned articles the team led by Dalakova and co-workers failed to prepare homogeneous samples of strontium-substituted lanthanum cuprate with an ultra-low concentration of Sr atoms ($x = 0.001$) and instead found evidences of the Mott Hubbard dielectric $La_2CuO_4$ as is shown in Table 1 and discussed in this paper.

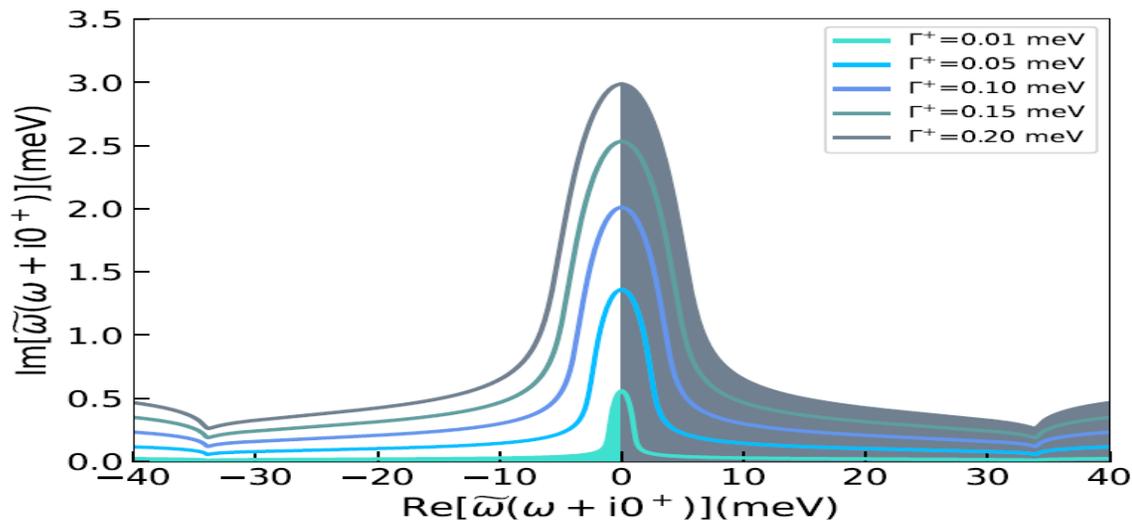

Fig. 4: The reduced scattering phase space RPS - ($\Re(\widetilde{\omega})$, $\Im(\widetilde{\omega})$) plot showing the different phases explained in Section 3, the imaginary part of the scattering cross-section as function of the non-magnetic impurities parameters $\Gamma^+$ is calculated for 5 values, a new sixth smallest value f $\Gamma^+$= 0.005 meV have been very recently in reference [36] .

## 4. Conclusions and recommendations.

In this work, we qualitatively explain experimental data and build a phenomenological model of $La_{2-x}Sr_xCuO_4$ for which three $x$-doping phases are distinguished as Dalakova and co-workers experimentally found on lightly strontium-substituted lanthanum cuprate samples. Using the idea of an energy concentration profile ($n\,(\widetilde{\omega}) \sim \Gamma^+/g(\widetilde{\omega})$) of normal state dressed quasiparticles based on the imaginary and real parts of SCCS, for different values of experimental and theoretical doping, a phenomenological energy density profile $n_{dressed}\,(\widetilde{\omega})$ can be zero, almost constant for most of the frequency interval, or strongly self-consistent and frequency-dependent in the whole energy (frequency) interval, depending on the impurity atoms concentration $\Gamma^+$, for the strong scattering unitary limit.

In there were distinguished: one almost antiferromagnetic phase where $n\,(\widetilde{\omega}) \to 0$, without dressed quasiparticles; another metallic coalescent lightly-substituted Sr phase with $n_{dressed}(\widetilde{\omega}) \to const$ with a small amount of non-magnetic impurities; and, finally, the strange metallic phase in dopped $La_{2-x}Sr_xCuO_4$, where the scattering lifetime is strong self-consistently dependent.

According to this model, as the antiferromagnetic-superconducting phases approach each other, the substitution of heavy lanthanum for lighter strontium atoms led to an increase in the frequency of vibrations near a single local Sr atom, thus significantly increasing the local temperature of the crystal lattice, therefore, at a particular concentration, the accumulation of strontium atoms in one place is energetically favorable, since it lowers the local temperature, which is the basis of coalescence phenomenon (Fig. 2). Thus, single non-magnetic strontium atoms tend to merge into metallic clusters, and, since the linear momentum of dressed quasiparticles $k$ is not conserved in the unitary metallic limit, it is transferred to strontium atoms in the crystal lattice, which migrate through the lattice and stick together forming the coalescing metallic cluster, for the lowest concentrations of non-magnetic strontium atoms.

This finding of the metallic coalescent phase with a constant scattering "$\tau_s$" agrees with a hypothesis firstly proposed by Schmitt-Rink, Miyake and Varma [29] to fit low temperatures transport superconducting data in unconventional superconductors. They proposed that the superconducting scattering lifetime "$\tau_s$" is almost constant and almost the same as in the normal metallic state "$\tau_{normal}$". Additionally, to this amendment, in the past, we were able to fit the low temperature superconducting data from directional superconducting ultrasound attenuation [30, 31] and electronic superconducting thermal transport [32, 33] experiments in clean samples of the triplet superconducting $Sr_2RuO_4$ using their proposal and a TB anisotropic model.

On the other hand, recently, using a similar method, we found three phases in the triplet compound strontium ruthenate, two phases with quasinodal points (one of them very inhomogeneous that corresponds to a tiny gap triplet model) and one phase with point nodes and the ground state similar to the HTSC line nodes model but with smaller $\Delta_0$, being all phases in the unitary scattering regime [34].

Therefore, we may conclude that the analysis of the elastic scattering cross-section in the RPS with a self-consistent procedure that accounts for the effects of the variations in the set of TB and disorder parameters ($\epsilon_F, t, \Delta_0, c, \Gamma^+$), is very instructive and deserves further attention (see Fig. 1 in Contreras, Osorio and Beliayev paper [35] for a general diagrammatic sketch used in this method). It serves as a recent proposed tool aiming at understanding some phenomenological aspects of unconventional superconductors, such as the different phases which depend on the impurity strength *c* and nonmagnetic disorder concentration $\Gamma^+$, but in the TB approach, extensions can be made to the flat band case [34-36]. We would like to clarify that some other aspects of Dalakova and coworkers work could be addressed in the future, following the ideas about the activation energy presented in the review by Prof. B. I. Beletsev [37], and also in the work by I. Lifshitz and collaborators, for example that for a system of low impurities concentration but strongly scattering impurities, the use of a one-site approximation and the first power in impurities concentration $\Gamma^+$ must be quite effective [38] as the formalism briefly presented in this work points out.

### Acknowledgments

The authors did not receive any financial support for this article's research, authorship, and/or publication. Experiments were performed by Drs. Dalakova and Beliayev and their co-workers at B. Verkin Institute for Low Temperature Physics and Engineering of the National Academy of Sciences of Ukraine. Numerical calculations are attributed to Profs. Contreras and Osorio, and the physical analysis and visualization of this particular problem belongs to the three participants of this research.